\documentclass[12pt,preprint]{aastex}

\shorttitle{Rest-frame $K$-band cluster galaxy luminosity function}
\shortauthors{De~Propris et al.}

\begin{document}


\title{The rest-frame $K$-band luminosity function of galaxies
        in clusters to $z=1.3$}


\author{Roberto~De~Propris\altaffilmark{1,2}, S. A. 
Stanford\altaffilmark{3,4}, Peter R. Eisenhardt\altaffilmark{5,6}, \\
Brad P. Holden\altaffilmark{7}, Piero Rosati\altaffilmark{8}}

\altaffiltext{1}{Cerro Tololo Inter-American Observatory, 
                 Casilla 603, La Serena, Chile}
\altaffiltext{2}{Department of Physics, University of Bristol,
                 Tyndall Avenue, Bristol, BS8 1TL, United Kingdom}
\altaffiltext{3}{Department of Physics, University of California at Davis, 
                 1 Shields Avenue, Davis, CA, 95616}
\altaffiltext{4}{Institute of Geophysics and Planetary Physics, 
                 Lawrence Livermore National Laboratories, L-413, 
                 Livermore, CA, 94550}
\altaffiltext{5}{MS 169-327, Jet Propulsion Laboratory, 4800 Oak
                 Grove Drive, Pasadena, CA, 91109}
\altaffiltext{6}{California Institute of Technology, 1200 E. California Boulevard,
Pasadena CA 91125}
\altaffiltext{7}{Lick Observatory, University of California, Santa Cruz, CA, 95604}
\altaffiltext{8}{European Southern Observatory, Karl Schwarzschild Stra{\ss}e 2,
                     85748 Garching, Germany}


\begin{abstract}

We derive the rest-frame $K$-band luminosity function for galaxies in 
32 clusters at $0.6 < z < 1.3$ using deep $3.6\mu$m and $4.5\mu$m imaging 
from the Spitzer Space Telescope InfraRed Array Camera (IRAC). The luminosity
functions approximate the stellar mass function of the cluster galaxies.
Their dependence on redshift indicates that massive cluster galaxies (to
the characteristic luminosity $M^*_K$) are fully assembled at least at 
$z \sim 1.3$ and that little significant accretion takes place at later times. 
The existence of massive, highly evolved galaxies at these epochs is likely 
to represent a significant challenge to theories of hierarchical structure 
formation where such objects are formed by the late accretion of spheroidal 
systems at $z < 1$.

\end{abstract}




\section{Introduction}

Clusters of galaxies are important for studies of galaxy formation and
evolution, because they contain a {\it volume limited} population of
galaxies observed {\it at the same cosmic epoch}. They therefore provide
a well-defined sample of objects to cosmologically significant lookback
times whose member galaxies can be identified  by simple counting statistics, 
without need for extensive spectroscopic surveys or multi-color data.

One important characteristic of early type galaxies in clusters is that
they are known to follow tight color-magnitude relations, which appear 
to be universal and to have very small intrinsic scatter to the highest 
redshifts yet observed \citep{visvanathan77,bower92,stanford95,
stanford98,blakeslee03,lopezcruz04,holden04,mei06a,mei06b}. Together 
with the conventional interpretation of the color-magnitude relation as 
a mass-metallicity correlation (e.g., \citealt{trager00}), this implies 
that the majority of the stellar populations in early-type cluster galaxies 
were formed via rapid dissipative starbursts at $z > 2$. Fundamental plane 
studies of high redshift cluster galaxies also support this conclusion, at 
least for the more massive objects \citep{vandokkum03,wuyts04,holden05}, 
although the low mass galaxies seem to have undergone more extended star
formation histories \citep{poggianti01,nelan05,jorgensen05}.

Theoretically, the existence of such massive and old galaxies at high redshift
should represent a severe challenge to models where galaxies are assembled
hierarchically, from a sequence of major mergers at progressively lower
redshifts (e.g., \citealt{coles05,springel05,baugh06} for recent reviews). 
It is not possible, however, to exclude by spectrophotometry alone, that these 
galaxies are assembled from sub-units whose star formation has already 
ceased, but which are not accreted until later times (similar to the
so-called `dry' mergers -- \citealt{vandokkum99,tran05}). This is assumed to
be the main channel by which spheroids grow at $z < 1$ in the hierarchical
scenario.

On the other hand, if galaxies are formed via mergers, we should observe
a steady decrease of the mean stellar mass in galaxies as we go to higher
lookback times and the most massive members of the merger tree branch into
ever smaller twigs \citep{delucia06,maulbetsch06}. While it is 
generally difficult to measure galaxy masses, the $K$-band luminosity 
function is believed to provide an adequate surrogate \citep{
kauffmann98}, as the rest frame $H$ or $K$ luminosity of galaxies is seen 
to correlate well with stellar and even dynamical mass for local samples 
\citep{gavazzi96,bell01} and even for high redshift galaxies \citep{
drory04,papovich05,caputi06}. 

In our previous work \citep{depropris99} we showed that the observed 
(ground-based) $K$-band luminosity of galaxies in clusters was consistent 
with pure passive evolution of objects formed at high redshift and argued 
that this implied that the majority of the stellar mass was completely 
assembled by at least $z=0.9$. More recent luminosity function studies have
essentially confirmed and extended this picture of early galaxy assembly 
in clusters (\citealt{kajisawa00,nakata01,massarotti03,kodama03}; Toft, 
Soucail \& Hjorth 2003; \citealt{ellis04,toft04,bremer06,lin06,strazzullo06}). 
\cite{andreon06} recently derived a composite 3.6$\mu$m luminosity function 
for galaxies in clusters in the XMM-LSS survey (at a mean redshift of 0.5), 
finding that the results are consistent with the previous ground-based 
results.

Ideally, we would wish to carry out this experiment in the {\it rest frame}
$K$-band, as even the ground-based $K$-band starts starts to sample the 
rest-frame optical at $z > 1$. The Spitzer Space Telescope \citep{werner04} 
InfraRed Array Camera \citep{fazio04a} is now capable of obtaining panoramic
($\sim 5' \times 5'$) images at $\lambda > 3\mu$m with $\mu$Jy sensitivity
and allows us to study the rest-frame $K$-band luminosity function of
cluster galaxies at high redshift.

Here, we present a study of 32 clusters up to $z=1.3$ in both the
$3.6\mu$m and $4.5\mu$m bands and derive the evolution of the rest-frame 
$K$-band galaxy luminosity function, which is a close proxy for the stellar mass 
function. We adopt the cosmological parameters $\Omega_M=0.3$, 
$\Omega_{\Lambda}=0.7$ and H$_0=70$ km s$^{-1}$ Mpc$^{-1}$.

\section{Data Reduction and Photometry}

The sample consists of 32 clusters at $0.6 < z < 1.3$. Data were obtained with
IRAC in all four filters, using 5 dithered frames of 200s each.  Here we discuss 
observations in the $3.6\mu$m and $4.5\mu$m filters, which map more closely 
to the rest-frame $K$-band at the redshifts of our clusters. 

Table 1 shows a list of clusters and some relevant properties. Most of the 
sample comes from the ROSAT Deep Cluster Survey (RDCS -- \citealt{rosati98}), 
while a few others derive from other X-ray or optical surveys (see table for
details), but the target selection is somewhat heterogeneous, especially for 
the higher redshift objects. On the other hand, \cite{depropris03a} and 
\cite{popesso05} have shown that the $B$-band galaxy luminosity function does
not depend on cluster properties such as the velocity dispersion, Bautz-Morgan 
type or richness and \cite{depropris99} found no difference in the ground-based 
$K$-band luminosity function of clusters selected by density and X-ray luminosity.

The IRAC data were reduced following standard procedures.   The raw (BCD)
data were first corrected for known IRAC artifacts associated with bright 
stars (mux-bleed and column pulldown). Then the SSC (Spitzer Science 
Centre) MOPEX software was used to mosaic the individual frames into a
registered mosaic, with cosmic rays removed.  This mosaiced image for each 
cluster and for each band has a pixel scale which is 1.414 times smaller 
than the original $1.2''$ IRAC pixel scale, and  the orientation rotated by 
$\sim 45^{\circ}$.

Photometry was carried out using the Source Extractor software \citep{
bertin96}. We experimented with various values for the background level
and the deblending threshold, because Spitzer images have relatively
poor angular resolution ($1.7''$ FWHM for stellar sources) and because 
our fields are moderately crowded. 

In order to deal with the moderate crowding, we checked that the poorer
resolution of Spitzer data does not significantly affect our detection and
photometry. We verified the detections visually, both on the original image 
and on the aperture image produced by Sextractor. We also compared photometry 
in $2''$, $3''$ and Kron apertures, extrapolated to total magnitudes, to 
check that objects were properly deblended. These tests provide confidence 
that our photometry is not significantly affected by the crowding, although
to fully address this issue will require higher resolution imaging.

\section{Number Counts for Cluster Galaxies}

We chose to measure magnitudes in fixed $3''$ (radius) apertures, which 
were calibrated on to the Vega system and extrapolated to infinity following 
\cite{fazio04b} to produce total magnitudes. This is done for consistency
with the apertures used by \cite{fazio04b} to derive galaxy number counts
in IRAC bands, which we use for background subtraction. We then counted objects
(stars and galaxies) in 0.5 magnitude wide bins within a circular aperture 
of radius 1 Mpc, centred on the brightest cluster galaxy (where the cluster
density is higher with respect to background). These systems tend to lie 
at or near the peak in galaxy density  and the centre of the X-ray isophotes. 
Ideally, we would wish to choose an aperture based on the cluster structural 
parameters (e.g., $r_{200}$) but the IRAC field size is not sufficiently 
large to derive a reliable profile for the cluster galaxy distribution. 
Since it is unlikely that the few bright galaxies at large cluster-centric 
radii may affect the luminosity function parameters significantly, and since 
there is little evidence that the luminosity function varies with distance 
from the cluster centre, at least for bright galaxies (\citealt{depropris03a}, 
Lin, Mohr \& Stanford 2004, \citealt{popesso05}), our choice of a 1 Mpc aperture 
should not affect our conclusions.

We estimated the contribution of background galaxies to the observed
counts by using the $3.6\mu$m and $4.5\mu$m counts of \cite{fazio04b}.
We fitted a low-order polynomial to the literature counts to smooth
the effects of large scale structure along the lines of sight of the
background fields. Errors were assumed to be Poissonian, while the 
clustering contribution was calculated following \cite{huang97} and
\cite{driver03}. The Poisson errors for the  cluster counts and the 
background contribution and the clustering errors for the field 
contribution  were then added in quadrature as appropriate.

Because of the low resolution of Spitzer data, we are not able to
discriminate easily between stars and galaxies. There are no
published star count models for Spitzer passbands. We estimated the
stellar contribution using the predicted $L$ band counts from the Besan{\c c}on
model of the Galaxy \citep{robin03}. These give a good fit to the star counts
reported by \cite{fazio04b} in the Extended Groth Strip and QSO 1700 fields.

Table 2 shows the raw number counts, estimated background contribution,
stellar contamination and corrected number of (statistical) cluster
members to the limiting apparent magnitude we use for both IRAC bands. 
The limiting magnitude is designed to reach the same absolute magnitude 
in all clusters (in two broad redshift ranges -- see discussion below), 
such that the cluster counts in the faintest bin are still significantly 
above the predicted contamination. At the same time, the brighter magnitude
limit (typically around 18$^{th}$ magnitude) reduces the effect of crowding,
which is most significant for the fainter galaxies. This limit is found to 
lie below the knee of the luminosity function and in the regime where the 
counts are fitted by a power law. 

There are some objects where the level of contamination 
from foreground stars is high or which have low number counts. These objects 
are identified in Table 2 and not used in our analysis. In practice, we choose
objects where the residual cluster counts (after removal of background galaxies 
and stars) are higher than 50 in the $3.6\mu$m band for the $ z < 0.9$ sample 
(where this band is closer to the rest-frame $K$) and larger than 25 for the 
$z > 1.1$ sample in the $4.5\mu$m band (i.e., where this passband better probes 
the rest-frame $K$) - Cl0848.9+4452 is an exception to this, as we have only 
1/4 of the field of other clusters. We remark that there is no evidence for the 
actual existence of our highest redshift target (QSO1215-00), suggesting that 
the structure identified by \cite{liu00} consists of a small group or filament.

The actual counts for each cluster suffer from small number statistics.
Rather than recovering the luminosity function from Bayesian methods (e.g.,
Andreon, Punzi \& Grado 2005) we use composite luminosity functions, 
in order to average errors out (c.f. \citealt{andreon06}). We create 
composite luminosity functions for clusters in two redshift bins, at $0.6 
< z  < 0.9$ and $1.1 < z < 1.3$, in both bands, following the procedure 
described by \cite{colless89}. We bin galaxies in 0.5 magnitudes wide absolute 
magnitude bins in rest-frame $K$, adopting the cosmology specified above and 
a $k$-correction derived using the models of \cite{bruzual03} to transform
from the observed IRAC bands to rest-frame $K$. We choose to sample these 
two redshift bins for the following reasons. Most previous studies, starting 
from \cite{depropris99} have studied clusters at $z < 1$; only recently have 
adequate cluster samples at $z > 1$ become available (e.g., \citealt{toft03,toft04,strazzullo06}). The two redshift bins we study sample the 
rest frame $K$-band luminosity function of galaxies in these two regimes, 
i.e. both the reasonably well studied $z < 1$ sample and the more recent 
clusters at higher redshift. Furthermore, the $3.6\ \mu$m band maps closely
to rest frame $K$ for the $0.6 < z < 0.9$ interval, while the $4.5\ \mu$m
band does the same for the $1.1 < z < 1.3$ regime.

The $k$-correction used above assumes a solar-metallicity single stellar
population formed at $z=3$ and with star formation declining exponentially
with an e-folding time {$\tau$) of 0.1 Gyr, and is computed independently
for each Spitzer band, which is thus transformed to rest-frame $K$. This is
done (rather than more complex approaches using both IRAC bands to derive the
rest-frame $K$ luminosity) to present the data more directly and with a 
minumum of model dependencies. The $k$-corrections used are presented in
Table 3 (the full version is available electronically, while we only
show the first ten lines of the table for guidance here). We experimented 
with several `reasonable' values of $\tau$ from instantaneous star formation 
to 1 Gyr e-folding time bursts and found that this makes little difference 
to the actual value of the $k$-correction.

The resulting composite luminosity functions are fitted with a Schechter
function, using a downhill simplex algorithm. Figure 1 shows the data in 
each bin and the best fitting luminosity functions. Table 4 shows the
derived $M_K^*$ values for the luminosity function in both bands. The
errors in $M^*$ are marginal 1$\sigma$ errors derived by fixing the
values of all other parameters at their `best' value. The derived $\alpha$
is also shown in Table 4, but we caution that the fit to the faint-end slope
is very uncertain.

\section{Discussion}

Figure 2 shows the rest-frame $K$ band $M^*$ for our composite clusters, 
together with previous ground-based $K$ band data, corrected to rest-frame
$K$ following the same procedure as above, and a \cite{bruzual03} model, 
with solar metallicity (Salpeter Initial Mass Function and Padova 1994
isochrones, as recommended by \citealt{bruzual03}) and variable formation epoch, 
with $\tau=0.1$ Gyr. The actual of choice of $\tau$ makes a difference only 
to the level of a few hundreds of magnitudes. Note that we are not attempting 
to actually fit models to the data shown in Fig. 2, but we are showing models 
with representative star formation histories to obtain upper limits to the epoch 
of mass assembly for these cluster galaxies.

The results shown in Fig. 2 imply that the majority of the stellar mass
in elliptical galaxies is already assembled at least at $z=1.3$: this is
a strong upper limit to the epoch of galaxy formation in that the majority
of the merger episodes (if any) must have taken place prior to this epoch.
Taken at face value, these Spitzer data may indicate that the epoch of star
formation in these objects is somewhat more recent ($1.5 < z < 2$) than indicated
by some previous studies of the color-magnitude relation \citep{stanford95,
stanford98,blakeslee03,holden04,mei06a,mei06b,holden06} and the fundamental 
plane (\citealt{wuyts04}, but see \citealt{jorgensen05}). This may be due to the 
fact that, at $\sim M^*$, the population includes a fraction of lenticular galaxies, 
whose star formation histories are more extended than the brighter elliptical galaxies. 
Otherwise, our results are consistent with recent work 
(e.g. \citealt{holden06}) showing that massive ellipticals are in place at 
$z \sim 1$ but are more general, as we make no morphological selection.

The latest hierarchical models (e.g., \citealt{delucia06}), including AGN feedback, 
succeed in pushing back the epoch of major star formation to $z > 2$ for
the most massive objects, but still require most of the actual mass assembly
to take place through dry mergers at later epochs; 50\% of the mass in 
more massive galaxies ($M > 10^{11}\ M_{\odot}$) is assembled at $z < 0.8$, 
while the lower mass objects ($M > 4 \times 10^9\ M_{\odot}$) may be formed 
at higher redshifts. This is not what we observe here, where the vast majority 
of the stellar mass in massive (approximately $L^*$ or greater, equivalent to
a mass of $\sim 10^{11.8}\ M_{\odot}$, using the calibration shown in \citealt{
gavazzi96}, which corresponds more closely to the virial mass - using stellar 
masses this is approximately $10^{11.3}\ M_{\odot}$) galaxies appears to be 
in place at $z=1.3$.

One possible caveat is that the theoretical models refer to the average
`field' environment, while we are observing massive clusters where the
main process of hierarchical merging and collapse may have taken 
place at earlier epochs, as they lie in overdense regions. However, 
\cite{maulbetsch06} use a high-resolution N-body simulation to simulate how 
the mass assembly histories of galaxy-size haloes depend on environment 
and show that at  $z=1$ the mass aggregation rate is 4 times higher than 
at present, and independent of environment, while galaxies in the densest 
(cluster-like) environments at $z > 1$ undergo more rapid mass accretion. 
This suggests that we should be witnessing a much stronger evolution of 
the mean galaxy mass than observed here, even for cluster environments. 

Our results are therefore largely inconsistent with the hierarchical
picture. Not only are the stellar populations of these galaxies formed
at high redshift (see Introduction), but they are also assembled into
galaxies at comparatively high lookback times, arguing that star formation
takes place {\it in situ}, in a manner reminiscent of the earlier monolithic 
collapse picture. Recently, it has been shown that this behavior largely 
holds for field ellipticals as well, at least to $z \sim 0.65$ 
\citep{roseboom06,wake06}. Similarly, $K$-selected studies in the
field have also found a similar anti-hierarchical behavior (e.g.,
\citealt{cimatti04} from the $K20$ survey -- see also the review
by \citealt{renzini06} for the observed `top down' buildup of
massive ellipticals as opposed to the theoretically favored
`bottom up' scenario).

Because the power spectrum of density fluctuations in the Universe at the 
time of recombination is tilted to low masses in the $\Lambda CDM$
scenario, hierarchical accretion is a necessary consequence of the hypothesis
that galaxies are formed within cold dark matter haloes. The evidence presented 
here poses a severe challenge to the hierarchical formation scenario 
in that the observations show the {\it opposite} behavior to the theoretical 
predictions, with the more massive galaxies being already present at
a lookback time of 65\% of the Hubble time.

{\sc Note added in proof}: D. Stern (priv. comm.) has obtained a deep {\it Chandra}
image of QSO1215-00 and finds no evidence of diffuse X-ray emission.

\section*{Acknowledgements}

We would like to acknowledge the anonymous referee for an informative
report that helped make the paper clearer. We also thank Steve Willner
for some comments on this paper. This work is based on observations made 
with the Spitzer Space Telescope, which is operated by the Jet Propulsion 
Laboratory, California Institute of Technology, under a contract with NASA.
RDP acknowledges support from a PPARC grant while at the University of
Bristol. SAS's work was performed under the auspices of the U. S.
Department of Energy, National Nuclear Security Administration 
by the University of California Lawrence Livermore National Laboratory
under contract No. W-7405-Eng-48.

\clearpage


%% If you are not including electonic art with your submission, you may
%% mark up your captions using the \figcaption command. See the 
%% No more than seven \figcaption commands are allowed per page, 
%% so if you have more than seven captions, insert a \clearpage 

\begin{figure}
\begin{tabular}{cc}
\includegraphics[width=2.65in]{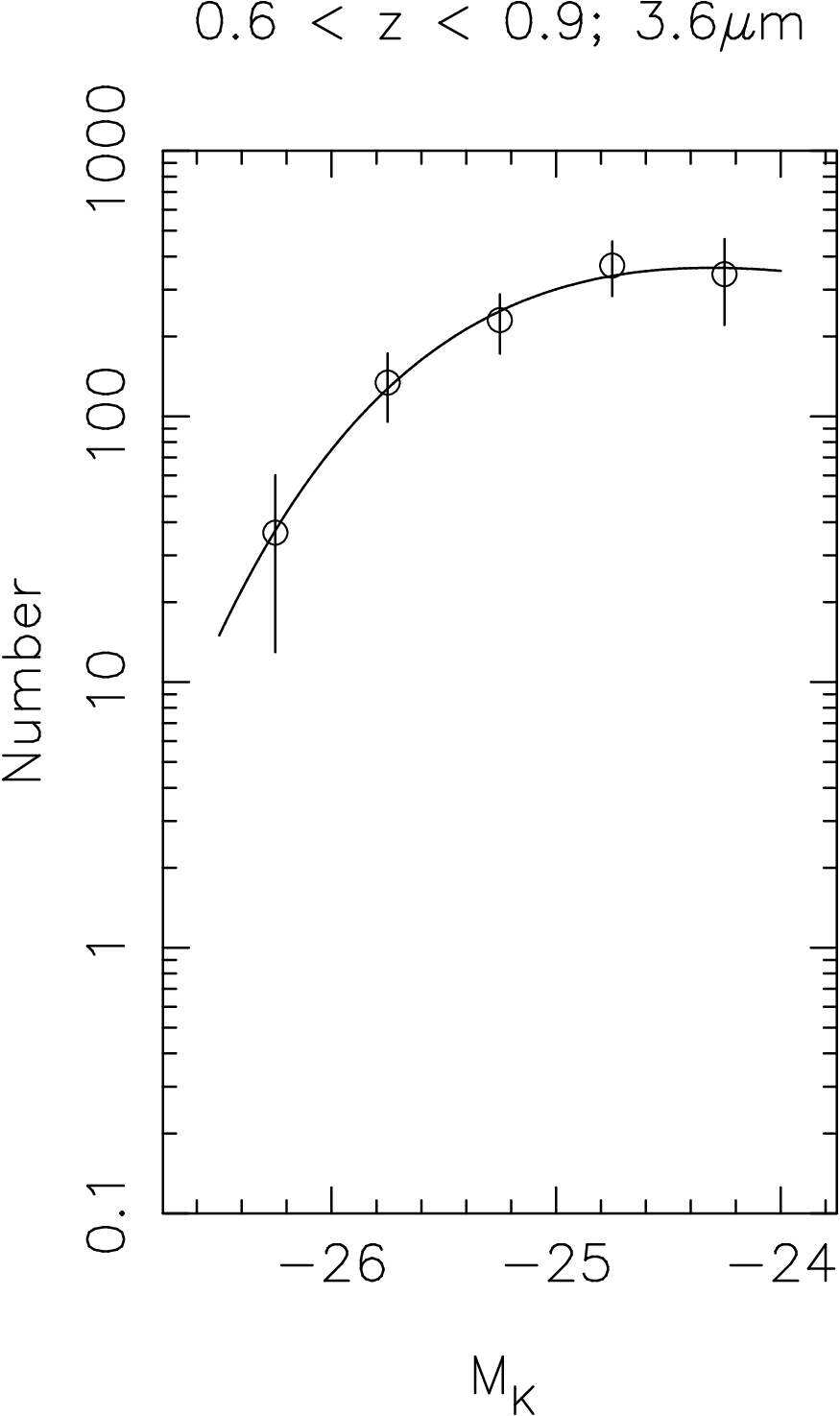} & \includegraphics[width=2.65in]{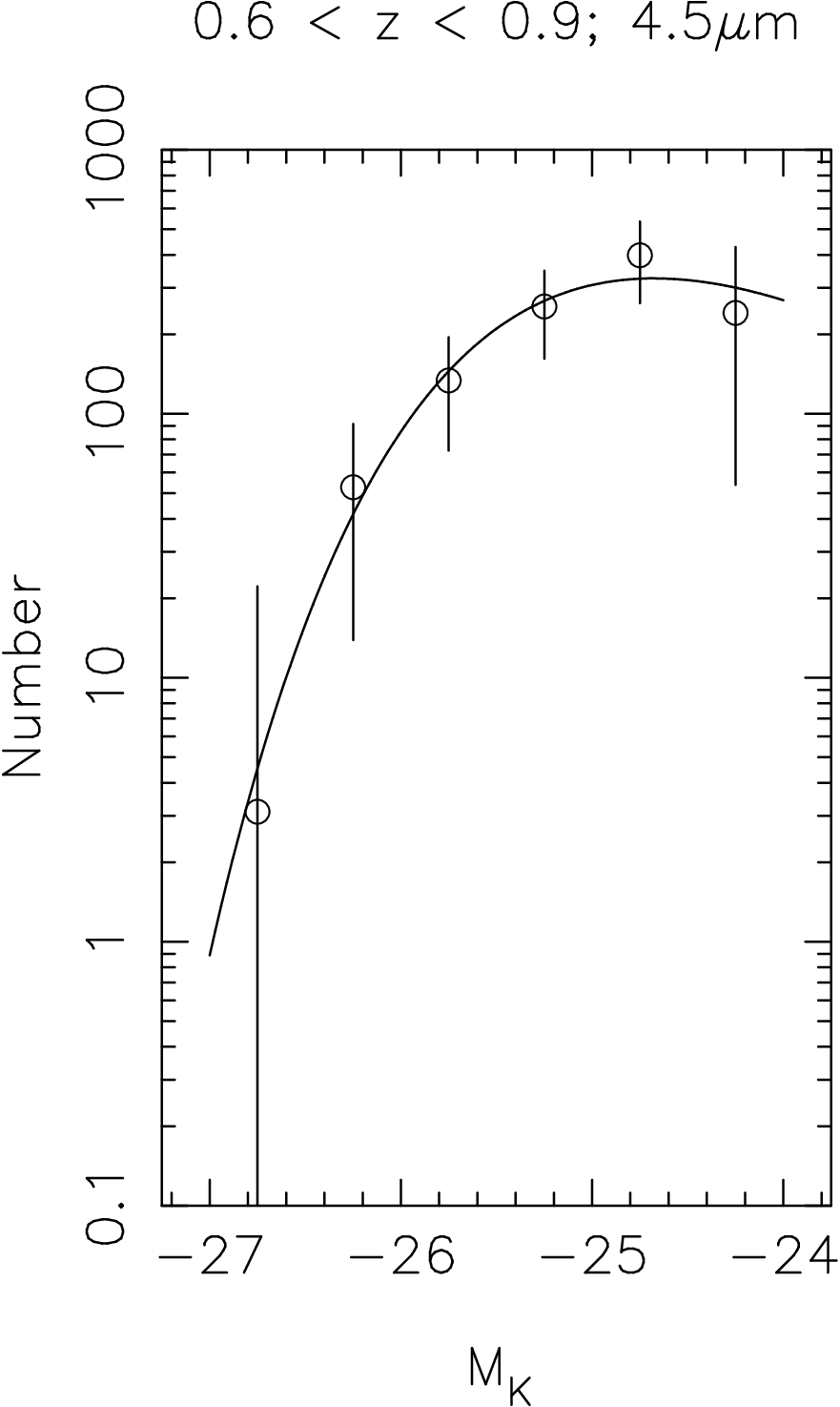}\\
 & \\
\includegraphics[width=2.65in]{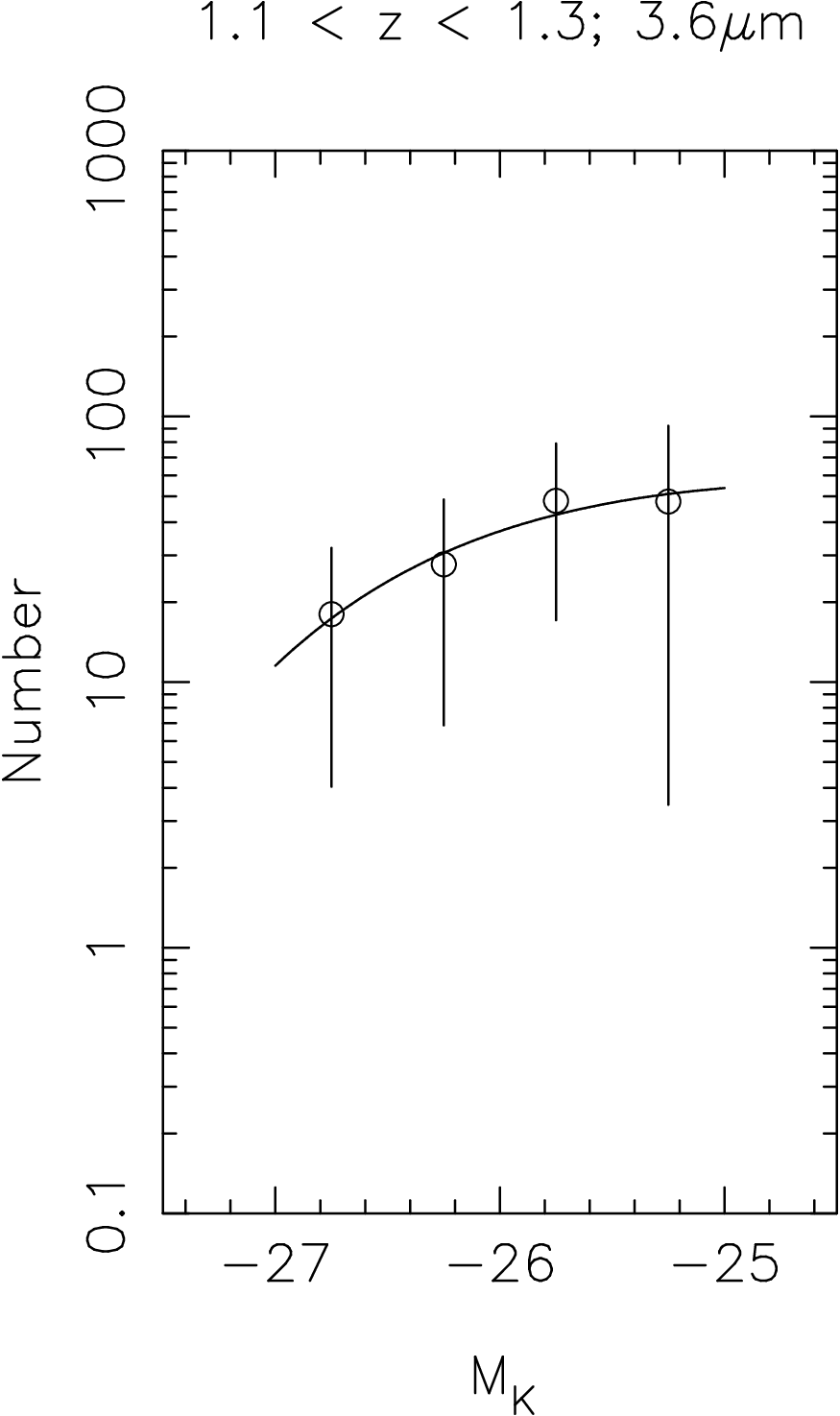} & \includegraphics[width=2.65in]{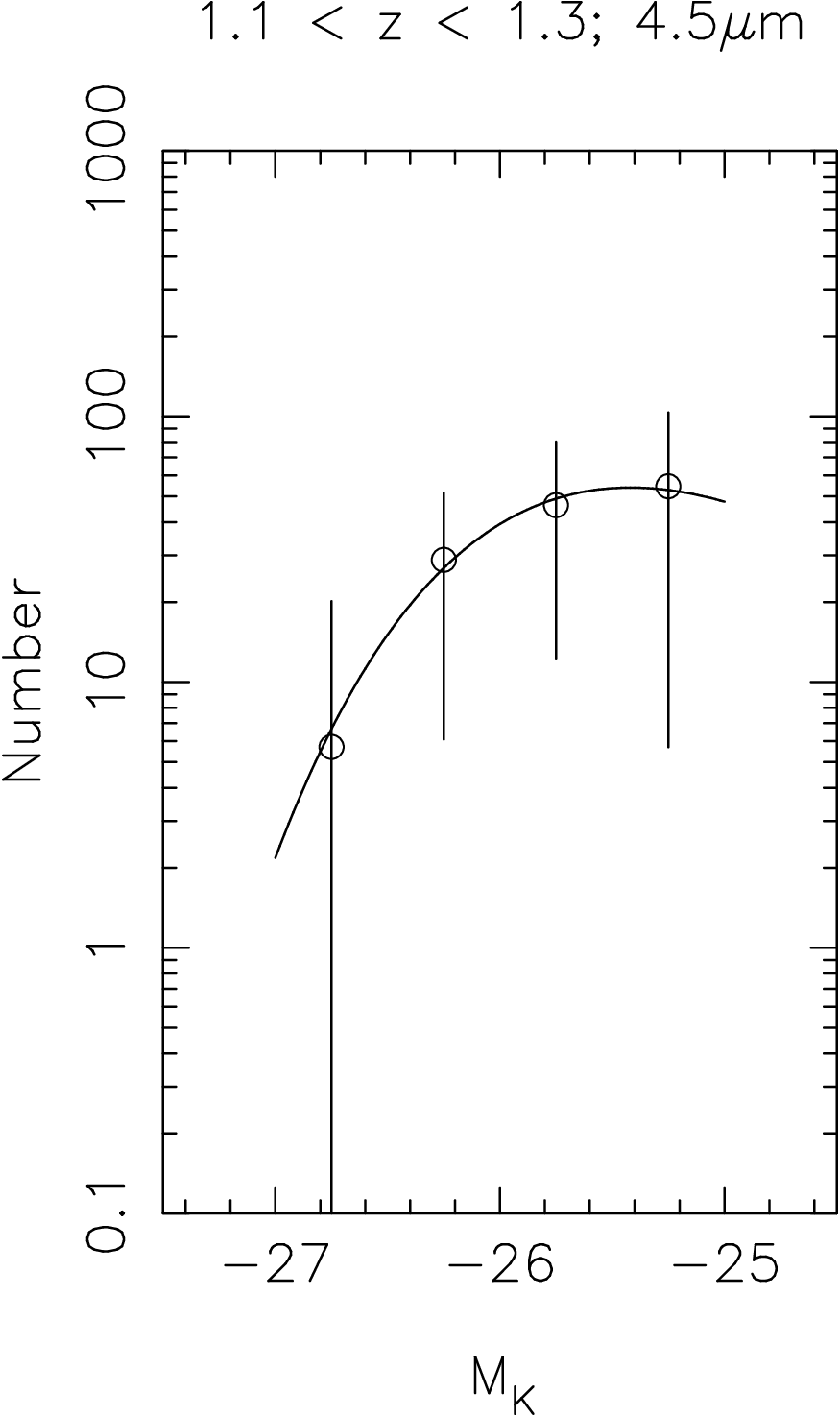}\\
\end{tabular}
\caption{Composite galaxy luminosity functions for cluster galaxies at $0.6 < z < 0.9$  and $1.1 < z < 1.3$
in Spitzer $3.6\mu$m and  $4.5\mu$m bands and best fits to a Schechter function.}
\end{figure}
\clearpage

\begin{figure}
\plotone{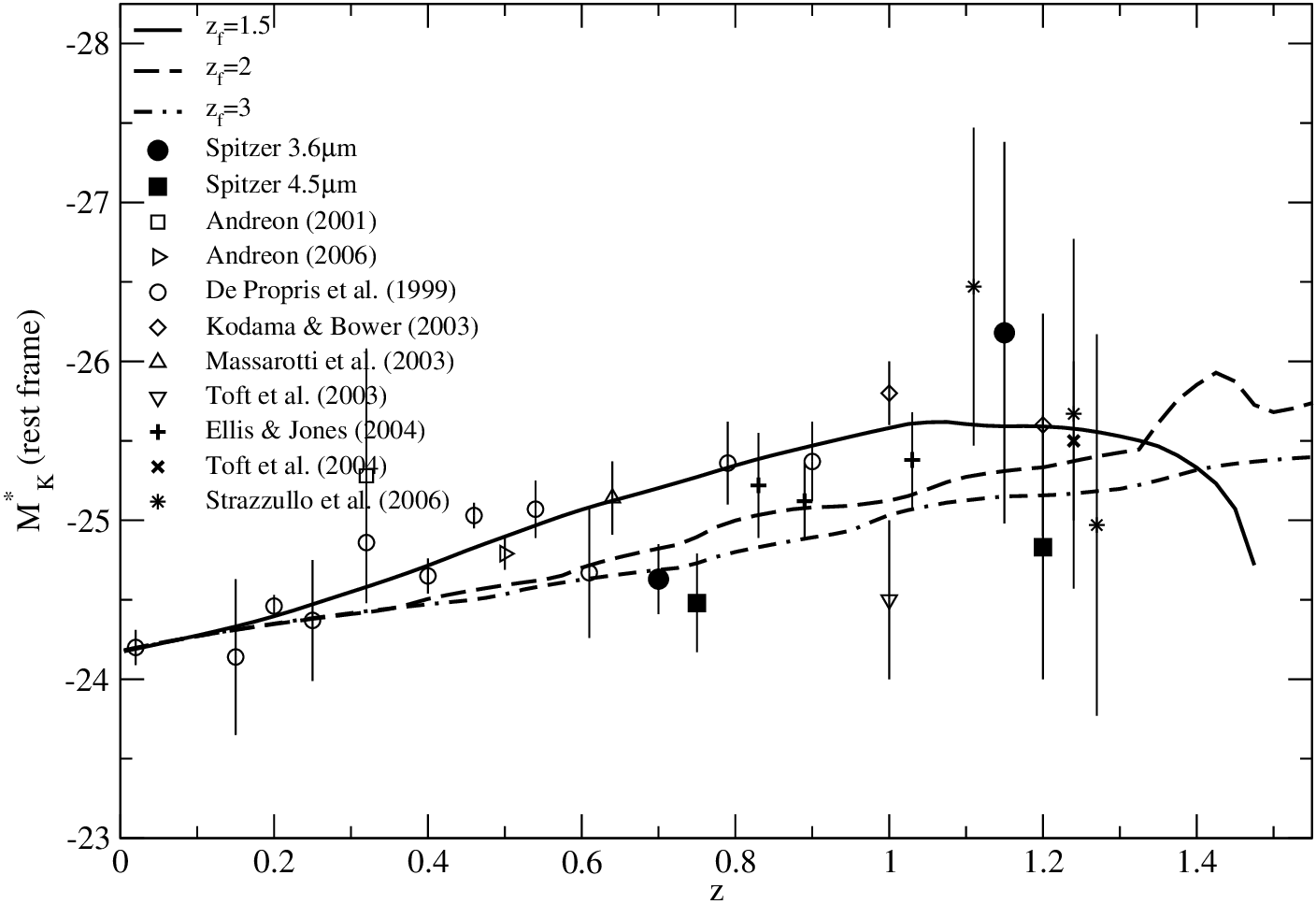}
\caption{Rest frame $M^*_K$ from our data (filled points) in the
$3.6\mu$m and $4.5\mu$m bands (arbitrarily shifted by $0.05$ in
$z$ for clarity), together with previous $K$-band studies (as
identified in the figure legend) and models from \cite{
bruzual03}, with $z_f$ as indicated in the legend and $\tau=0.1$
Gyr. The models are normalized to the value of $K^*$
in the Coma cluster \cite{depropris98}. }
\end{figure}

\clearpage

\thispagestyle{empty}

%% Tables should be submitted one per page, so put a \clearpage before

%% Two options are available to the author for producing tables:  the
%% deluxetable environment provided by the AASTeX package or the LaTeX
%% table environment.  Use of deluxetable is preferred.
%%

%% Three table samples follow, two marked up in the deluxetable environment,
%% one marked up as a LaTeX table.

%% In this first example, note that the \tabletypesize{}
%% command has been used to reduce the font size of the table.
%% Note also that the \label command needs to be placed 
\begin{deluxetable}{lcccl}
\tablewidth{0pt}
\tabletypesize{\footnotesize}
\tablecaption{Observed clusters}
\tablehead{
\colhead{Cluster} & \colhead{RA (J2000)} & \colhead{Dec. (J2000)} 
& \colhead{Redshift} & \colhead{Reference}}
\startdata
RJ1120+43 & 11:20:07.5 & +43:18:05.0 & 0.60 & Romer et al. 2000 \\ 
RDCS1634.5+5724 & 16:34:27.6 & +57:22:51.8 & 0.61 & Rosati et al. 1998\\ 
RDCS0046.3+8531 & 00:46:19.7 & +85:31:01.3 & 0.62 & "\\ 
RDCS0440.5-1630 & 04:40:28.4 & $-16$:30:08.0 & 0.62  & "\\
RDCS0542.8-4100 & 05:42:50.2 & $-41$:00:07.0 & 0.64  & "\\
MS1610+66 & 16:10:47.8 & +66:08:41.0 & 0.65 & Gioia et al. 1990 \\
RDCS1936.0-4640 & 19:36:06.6 & $-46$:40:03.6 & 0.65 & Rosati et al. 1998\\
RDCS0047.3+8506 & 00:47:14.8 & +85:06:02.0 & 0.66 & " \\
RDCS2313.6+1415 & 23:13:34.5 & +14:15:15.5 & 0.67 & " \\
RDCS2038.5-0125 & 20:38:29.1 & $-01$:25:11.7 & 0.68 & " \\
RDCS2236.7-2609 & 22:36:42.7 & $-26$:09:30.0 & 0.70 & " \\
RDCS2303.7+0846 & 23:02:47.5 & +08:44:07.4 & 0.73 & " \\
GHO1322+30 & 13:24:48.2 & +30:11:14.0 & 0.75 & Gunn, Hoessel \& Oke 1986 \\
RDCS1517.9+3127 & 15:17:56.3 & +31:27:27.0 & 0.75 & Rosati et al. 1998 \\
MS1137+66 & 11:40:22.3 & +66:08:15.0 & 0.78 & Gioia et al. 1990 \\
RDCS1350.8+6007 & 13:50:46.1 & +60:07:09.5 & 0.80 & Rosati et al. 1998 \\
RDCS0035.9+8513 & 00:35:55.2 & +85:13:20.0 & 0.81 & " \\
RDCS1317.4+2911 & 13:17:21.4 & +29:11:25.0 & 0.81 & "\\
RJ1716+67 & 17:16:49.6 & +67:08:30.0 & 0.81 & Henry et al. 1997 \\
RDCS0152.7-1357 & 01:52:43.7 & $-13$:57:21.0 & 0.83 & Rosati et al. 1998 \\
RDCS0337.4-3457 & 03:37:24.7 & $-34$:57:29.0 & 0.84 & " \\
RJ1226+33 & 12:26:54.0 & +33:32:00.0 & 0.89 & Ebeling et al. 2001 \\
GHO1604+4304 & 16:04:23.2 & +43:04:44.0 & 0.90 & Gunn, Hoessel \& Oke 1986 \\
3C184 & 07:39:24.5 & +70:23:11.0 & 1.00  & Deltorn et al. 1997 \\
MG 2019.3+1127 & 20:19:18.0 & +11:27:10.0 & 1.00 & Hattori et al. 1997 \\
RDCS0910.7+5422 & 09:10:45.0 & +54:22:02.0 & 1.11 & Rosati et al. 1998 \\
3C210 & 08:58:09.9 & +27:50:52.0 & 1.16 & J.-M. Deltorn, priv. comm. \\
3C324 & 15:49:48.9 & +21:25:38.0 & 1.21 & Dickinson 1995 \\
RDCS1252.9-2927 & 12:52:54.2 & $-29$:27:07.0 & 1.24 & Rosati et al. 1998 \\
RDCS0848.9+4452 & 08:48:56.2 & +44:52:00.0 & 1.26 & " \\
RDCS0848.6+4453 & 08:48:34.2 & +44:53:35.0 & 1.27 & " \\
QSO1215-00 & 12:15:49.8 & -00:34:34.0 & 1.31 & Liu et al. 2000 \\
\enddata
\end{deluxetable}
\clearpage

\begin{deluxetable}{lcccccccccc}
\tablewidth{0pt}
\tabletypesize{\footnotesize}
\setlength{\tabcolsep}{0.02in} 
\tablecaption{Number counts in clusters}
\tablehead{
\colhead{Cluster} & \colhead{$M_{lim}$} & \colhead{$N_{total}$} & \colhead{$N_{background}$} 
& \colhead{$N_{stars}$} & \colhead{$N_{cluster}$} & \colhead{$M_{lim}$} & \colhead{$N_{total}$} 
& \colhead{$N_{background}$} & \colhead{$N_{stars}$} & \colhead {$N_{cluster}$}\\
\colhead{} & \colhead{3.6$\mu$m} & \colhead{3.6$\mu$m} & \colhead{3.6$\mu$m} & \colhead{3.6$\mu$m}
& \colhead{3.6$\mu$m} & \colhead{4.5$\mu$m} & \colhead{4.5$\mu$m} & \colhead{4.5$\mu$m} & \colhead{4.5$\mu$m} 
& \colhead{4.5$\mu$m}}
\startdata
RJ1120+43 & 17.34 & 120 & 34.6 & 6.4 & 80.0 & 17.32 & 168 & 71.4 & 6.4 & 90.2 \\ 
CL1634+57$^a$ & 17.36 & 75 & 48.0 & 14.4 & 12.6 & 17.37 & 112 & 74.8 & 22.5 & 14.7 \\ 
CL0046+85$^a$ & 17.37 & 131 & 52.4 & 35.5 & 43.1 & 17.40 & 177 & 94.6 & 39.0 & 43.4 \\ 
CL0440-16 & 17.23 & 115 & 47.8 & 15.9 & 51.3 & 17.40 & 138 & 72.0 & 16.5 & 49.5 \\
CL0542-41 & 17.40 & 161 & 47.3 & 26.5 & 87.2 & 17.43 & 189 & 74.7 & 22.3 & 92.0 \\
MS1610+66 & 17.45 & 121 & 49.1 & 12.3 & 59.6 & 17.44 & 170 & 73.2 & 14.2 & 82.6 \\
CL1936-46$^a$ & 17.45 & 134 & 53.0 & 89.3 & $-8.3$ & 17.44 & 164 & 75.1 & 89.2 & $-0.3$ \\
CL0047+85$^a$ & 17.49 & 133 & 53.0 & 38.2 & 44.2 & 17.53 & 147 & 75.1 & 29.3 & 42.6 \\
CL2313+14 & 17.53 & 123 & 58.5 & 14.6 & 49.9 & 17.54 & 148 & 84.8 & 16.5 & 46.8 \\
CL2038-01 & 17.57 & 183 & 58.9 & 14.7 & 109.4 & 17.58 & 191 & 87.4 & 16.6 & 87.0 \\
CL2236-26 & 17.64 & 121 & 58.7 & 10.4 & 51.9 & 17.62 & 150 & 88.4 & 16.4 & 50.0 \\
CL2303+08$^a$ & 17.27 & 73 & 24.0 & 12.8 & 36.2 & 17.69 & 126 & 88.0 & 20.7 & 25.3 \\
CL1322+30 & 17.83 & 134 & 74.0 & 5.2 & 54.7 & 17.71 & 157 & 92.2 & 2.2 & 62.6 \\
CL1517+31$^a$ & 17.67 & 105 & 59.0 & 9.7 & 36.3 & 17.71 & 117 & 69.4 & 9.9 & 37.7 \\
MS1137+66 & 17.74 & 126 & 63.7 & 7.1 & 55.2 & 17.77 & 167 & 94.0 & 7.9 & 65.1 \\
CL1350+60 & 17.78 & 139 & 63.7 & 6.3 & 69.0 & 17.79 & 181 & 95.3 & 7.0 & 78.7 \\
CL0035+85 & 17.81 & 154 & 64.4 & 28.2 & 61.4 & 17.82 & 177 & 96.3 & 30.9 & 49.8 \\
CL1317+29 & 17.81 & 114 & 64.7 & 5.3 & 44.0 & 17.82 & 139 & 95.9 & 5.2 & 37.9 \\
RJ1716+67 & 17.81 & 167 & 64.4 & 19.8 & 82.8 & 17.82 & 184 & 97.3 & 22.2 & 64.5 \\
CL0152-13 & 17.85 & 161 & 67.1 & 6.0 & 87.9 & 17.82 & 189 & 97.5 & 6.8 & 84.7 \\
CL0337-34$^a$ & 17.88 & 106 & 67.7 & 12.8 & 25.5 & 17.84 & 144 & 97.0 & 14.6 & 32.4 \\
RJ1226+33 & 17.97 & 129 & 70.5 & 4.7 & 53.8 & 17.85 & 143 & 93.5 & 5.2 & 44.3 \\
GHO1604+4304$^a$ & 17.97 & 112 & 69.9 & 11.6 & 30.5 & 17.91 & 132 & 96.2 & 11.8 & 24.0 \\
3C184$^a$ & 17.16 & 72 & 57.7 & 11.2 & 3.1 & 17.02 & 51 & 34.6 & 11.3 & 5.1 \\
AXJ2019.3+1127$^a$ & 17.16 & 146 & 69.8 & 75.6 & 0.6 & 17.02 & 134 & 35.0 & 98.0 & 1.0 \\
CL0910+54 & 17.37 & 59 & 33.7 & 5.8 & 19.5 & 17.12 & 72 & 38.7 & 5.1 & 28.2 \\
3C210 & 17.32 & 62 & 29.1 & 20.7 & 12.2 & 17.18 & 74 & 40.7 & 6.1 & 27.2 \\
3C324 & 17.43 & 71 & 35.1 & 8.5 & 27.4 & 17.24 & 73 & 41.6 & 8.7 & 22.7 \\
CL1252-29 & 17.41 & 102 & 35.1 & 15.8 & 39.0 & 17.27 & 93 & 45.9 & 16.0 & 37.1 \\
CL0848.9+4452$^b$ & 17.54 & 39 & 8.6 & 1.7 & 34.7 & 17.30 & 21 & 11.1 & 1.7 & 9.3\\
CL0848.6+4453 & 17.54 & 69 & 34.4 & 6.5 & 28.1 & 17.30 & 79 & 46.3 & 6.7 & 26 \\
QSO1215-00\tablenotemark{a} & 17.75 & 55 & 46.2 & 4.8 & 4.0 & 17.36 & 48 & 45.4 & 4.4 & $-1.8$ \\
\enddata
\tablenotetext{a}{\ Low number counts}
\tablenotetext{b}{\ Only 0.5 Mpc field - cluster is off centre}
\end{deluxetable}
\clearpage

\begin{deluxetable}{lcc}
\tablewidth{0pt}
\tablecaption{$k-$corrections}
\tablehead{
\colhead{Redshift} & \colhead{$k_{3.6}$} & \colhead{$k_{4.5}$}}
\startdata
0.01  & $-0.216$ & $-0.216$ \\
0.025 & $-0.249$ & $-0.268$ \\
0.05  & $-0.3  $ & $-0.352$ \\
0.075 & $-0.347$ & $-0.433$ \\
0.1   & $-0.391$ & $-0.512$ \\
0.125 & $-0.434$ & $-0.588$ \\
0.15  & $-0.475$ & $-0.66 $ \\
0.175 & $-0.516$ & $-0.724$ \\
0.2   & $-0.555$ & $-0.779$ \\
0.225 & $-0.594$ & $-0.833$ \\
\enddata
\end{deluxetable}
\clearpage

\begin{deluxetable}{lcccc}
\tablewidth{0pt}
\tablecaption{Luminosity Function Parameters}
\tablehead{
\colhead{Redshift} & \colhead{$M^*$ ($3.6\mu$m)} & \colhead{$\alpha$} 
& \colhead{$M^*$ ($4.5\mu$m) } & \colhead{$\alpha$}}
\startdata
0.75 & $-24.63 \pm 0.22$ & $-0.25$ & $-24.48 \pm 0.31$ & $0.21$ \\
1.15 & $-26.18 \pm 1.20$ & $-0.82$ & $-24.83 \pm 0.83$ & $0.81$ \\
\enddata
\end{deluxetable}
\clearpage

\end{document}